\documentstyle[prl,aps,epsf]{revtex}
\begin{document}

\twocolumn[
\hsize\textwidth\columnwidth\hsize\csname@twocolumnfalse\endcsname

\title{Quadratic Quantum Measurements}

\author{Wenjin Mao and Dmitri V. Averin}

\address{Department of Physics and Astronomy, Stony Brook
University, SUNY, Stony Brook, NY 11794-3800}

\author{Rusko Ruskov\footnote{On leave from
INRNE, Sofia BG-1784, Bulgaria} and Alexander N. Korotkov}

\address{Department of Electrical Engineering, University of
California, Riverside, CA 92521-0204}

\date{\today}

\maketitle

\begin{abstract}
We develop a theory of quadratic quantum measurements by a
mesoscopic detector. It is shown that quadratic measurements
should have non-trivial quantum information properties, providing,
for instance, a simple way of entangling two non-interacting qubits.
We also calculate output spectrum of a quantum detector with both
linear and quadratic response continuously monitoring coherent
oscillations in two qubits.

\end{abstract}

\pacs{PACS numbers: 03.65.Ud; 03.67.Lx; 73.23.-b}

]

The problem of quantum measurements with mesoscopic solid-state
detectors attracts considerable current interest (see, e.g.,
chapters on quantum measurements in \cite{b1}). This interest is
motivated in part by important role of measurement in quantum
computing, and in part by the possibility, provided by the
mesoscopic structures, to study directly both in theory and
experiment transition between quantum and classical behavior in
systems that are large on atomic scale. Although mesoscopic
detectors can be quite different and include, e.g., quantum
point contacts (QPC) \cite{q1,q2,q3,q4,q5,q6,q7}, normal and
superconducting SET transistors \cite{s1,s2,s3,s4,s5,s6,s7},
SQUID magnetometers \cite{b2} and generic mesoscopic conductors
\cite{m1,m2}, the operating principle of almost all of them is
the same. Measured quantum system controls
transmission amplitude $t$ of some particles (electrons, Cooper
pairs, or magnetic flux quanta) between the two reservoirs, so that
the flux of these particles provides information on the state of
the system \cite{b5}. In a generic situation the amplitude $t$ varies
together with some control operator $x$, and for sufficiently weak
detector-system coupling, the dependence $t(x)$ can be approximated
as linear. Dynamics of such linear measurements is well understood
(see, e.g., \cite{dva,ank}).

At some special bias points, however, the linear response
coefficient of the $t(x)$ dependence vanishes and this
dependence becomes quadratic. This can happen, for instance,
if the amplitude $t$ is formed by more than one interfering
tunneling trajectories. Known examples of such situation
include dc SQUIDs and superconducting SET transistors. In this
work, we show that quantum detector operating at such a special
point should enable measurements of product operators referring
to separate systems and have non-trivial quantum information
processing properties, e.g.,
create simple entanglement mechanism for non-interacting qubits.
Specifically, we consider the measurement of two qubits (Fig.\ 1)
which is the simplest system that reveals non-trivial
characteristics of quadratic detection. (Quadratic detectors
can not measure individual qubits since dynamic variables of one
qubit are given by the Pauli matrices $\sigma$ for which
$\sigma^2=1$.) The two qubits (indexed by $j=1,2$) are assumed
to be coupled to one detector through their basis-forming
variables, i.e., $x=c_1\sigma_z^1+c_2\sigma_z^2$ so that
\begin{equation}
t(x) = t_0+\sum_j \delta_j \sigma_z^{j}+ \lambda \sigma_z^{1}
\sigma_z^{2} \, .
\label{e4} \end{equation}
The last term in this equation appears due to non-linearity of
$t(x)$. If $\delta_j=0, \, \lambda \neq 0$, one has purely
quadratic detector.
For the two qubits, Eq.~(\ref{e4}) represents the most general
dependence of $t$ on $\sigma_z^{j}$, whereas for measurements of
other systems, expansion in the measured operators similar to
Eq.~(\ref{e4}) can be justified as Taylor's expansion in weak
detector-system coupling.

\begin{figure}[htb]
\setlength{\unitlength}{1.0in}
\begin{picture}(3.,1.2)
\put(.3,.1){\epsfxsize=2.4in\epsfbox{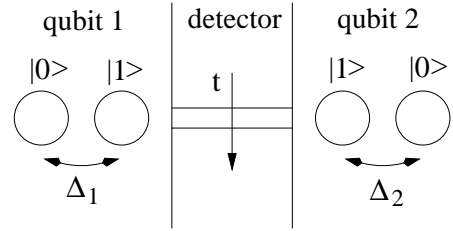}}
\end{picture}
\caption{Diagram of a mesoscopic detector measuring two qubits.
The qubits modulate amplitude $t$ of tunneling of detector
particles between the two reservoirs.}
\label{fig1} \end{figure}

The Hamiltonian of the detector-qubit system is:
\begin{equation}
H_t=H_0 + H_d + t(\{ \sigma_z^{j} \})\xi + t^{\dagger}
(\{ \sigma_z^{j} \}) \xi^{\dagger} \, ,
\label{e1} \end{equation}
where $H_0 = -(1/2) \sum_{j=1,2} (\varepsilon_j
\sigma_{z}^{j} + \Delta_j \sigma_x^{j}) + (\nu/2)
\sigma_{z}^{1} \sigma_{z}^{2}$.
Here $\Delta_j$ is the tunnel amplitude and $\varepsilon_j$
is the bias of the $j$th qubit, $\nu$ is the qubit interaction
energy, $H_d$ is the detector Hamiltonian, and $\xi^{\dagger},
\xi$ are the detector operators that create excitations when a
particle is transferred, respectively, forward and backward
between the detector reservoirs. For instance, for the QPC
detector, $\xi^{\dagger} , \xi$ describe excitation of
electron-hole pairs in the QPC electrodes. The qubit interaction
is not essential to us, and we will frequently take $\nu=0$
below.

We make two assumptions about the detector: the tunneling
between reservoirs is weak, so that the evolution of the system
can be described in the lowest non-vanishing order in $t$;
characteristic time scale of tunneling is much shorter than that
of the qubit evolution due to $H_0$. In the example of the QPC
detector these assumptions mean that the QPC operates in the
tunneling regime and that the voltage across it is much larger
than the qubit energies. Under these assumptions, precise form of
the detector Hamiltonian $H_d$ is not important and dynamics of
measurement is defined by the correlators
\begin{equation}
\gamma_+=\int_{0}^{\infty}dt\langle \xi(t) \xi^{\dagger}\rangle \, ,
\;\;\; \gamma_-=\int_{0}^{\infty}dt\langle \xi^{\dagger}(t)
\xi\rangle \, .
\label{e3} \end{equation}
In Eq.~(\ref{e3}),  the angled brackets denote averaging over
internal degrees of freedom of the detector reservoirs which
are taken to be in a stationary state. The correlators $\langle
\xi(t) \xi\rangle$, $\langle \xi^{\dagger}(t) \xi^{\dagger} \rangle$
that do not conserve the number of particles are assumed to vanish.

The measurement contribution to the evolution of the qubit
density matrix $\rho$ is obtained by standard perturbation theory
in tunneling and can be written down conveniently in the
``measurement'' basis of eigenstates of the $\sigma_z^{j}$
operators,  $|\! \uparrow \uparrow \rangle \, , |\! \uparrow
\downarrow \rangle \, ,|\! \downarrow \uparrow \rangle \, ,$ and
$|\! \downarrow \downarrow \rangle$. Each state $|k\rangle$ of
this basis is characterized by the value $t_k$ of the
transmission amplitude (\ref{e4}): $t_1=t_0+\sum_j \delta_j +
\lambda$, $t_2=t_0+\delta_1-\delta_2- \lambda$, $t_3=t_0-\delta_1
+\delta_2- \lambda$, $t_4=t_0-\sum_j \delta_j + \lambda$. To
describe qubit dynamics conditioned on particular outcome of
measurement, we keep in the evolution equation the number of
particles $n$ transferred through the detector. Since the
correlators that do not conserve $n$ vanish, only terms diagonal
in $n$ contribute to the evolution, and in the lowest order in
tunneling, measurement contribution to $\dot{\rho}$ is:
\begin{eqnarray}
\dot{\rho}^{n}_{kl}= -(1/2)(\Gamma_+ +\Gamma_-)(|t_k|^2+
|t_l|^2) \rho^{n}_{kl} +  \nonumber \\
\Gamma_-t_kt_l^*\rho^{n+1}_{kl}+ \Gamma_+ t_k^*t_l \rho^{n-1}_{kl}
- i[\delta H, \rho^{n}]_{kl} \, .
\label{e5} \end{eqnarray}
Here $\delta H=\sum_j \delta \varepsilon_j\sigma_z^{j}+ \delta \nu
\sigma_z^{1}\sigma_z^{2}$ is the renormalization  of the qubit
Hamiltonian due to coupling to the detector: $\delta \varepsilon_j
= \mbox{Re} (\delta_jt_0^* + \delta_{j'}\lambda^*) \mbox{Im}
(\gamma_- +\gamma_+)$ and $ \delta \nu = \mbox{Re} (\delta_1\delta_2^*
+ t_0\lambda^*) \mbox{Im} (\gamma_- +\gamma_+)$, where $j,j'=1,2$,
$j'\neq j$. Equation (\ref{e5}) is the basis for our quantitative
discussion of quadratic measurements. It generalizes to arbitrary
detector and two qubits, the equation obtained in \cite{q4} for
a qubit measured with the QPC in the tunnel regime.

Disregarding the index $n$ in Eq.\ (\ref{e5}), we obtain
equation for the measurement-induced evolution of the qubit
density matrix averaged over different measurement outcomes.
Together with the evolution due to the qubit Hamiltonian $H_0$
this equation is:
\begin{equation}
\dot{\rho}_{kl}= -\gamma_{kl}\rho_{kl} -i[H_0,\rho]_{kl}\, .
\label{e6} \end{equation}
Here $\gamma_{kl} \equiv (1/2)(\Gamma_+ +\Gamma_-)|t_k-t_l|^2$,
with $\Gamma_{\pm}\equiv 2\mbox{Re} \gamma_{\pm}$,
and we included in $H_0$ two renormalization terms: $\delta H$
(\ref{e5}) and $\delta H'$ due to phases $\varphi_{kl} \equiv
\arg(t_kt_l^*)$ of the transfer amplitudes in Eq.\ (\ref{e5})
defined by:  $[\delta H', \rho]_{kl}= (\Gamma_+ -\Gamma_-)|t_kt_l|
\sin \varphi_{kl}\rho_{kl}$.

Evolution (\ref{e6}) of the qubit density matrix is
reflected in the detector current. The form of the current $I$
operator in the qubit space is obtained by the same lowest-order
perturbation theory in tunneling that leads to Eq.\ (\ref{e6}):
\begin{equation}
I= (\Gamma_+ -\Gamma_-)t^{\dagger}t\, .
\label{ea1} \end{equation}
This equation can be used to calculate both the dc current
$\langle I\rangle =\mbox{Tr}\{I\rho_0\}$, where $\rho_0$ is the
stationary solution of Eq.\ (\ref{e6}), and the current spectral
density
\begin{equation}
S_I = S_0 + 2\int_0^{\infty} d \tau \cos\omega \tau
(\mbox{Tr} \{ I e^{L\tau} [ I\rho_0] \} - \langle I\rangle^2).
\label{ea2} \end{equation}
Here $S_0 = (\Gamma_+ +\Gamma_-)\mbox{Tr}\{ t^{\dagger}t\rho_0 \}$,
and $e^{L\tau} [A]$ denotes the evolution of the matrix A during
time interval $\tau$ governed by Eq.\ (\ref{e6}).

Decay of the off-diagonal matrix elements Eq.\ (\ref{e6})
is the result of averaging over the measurement outcomes. However,
since $n$ is the classical detector output, it is legitimate to ask
a question what is the qubit evolution for a specific measurement
outcome $n$. Such a ``conditional'' description of the measurement
dynamics (see, e.g., \cite{ank}) is convenient for calculation of
more complicated
correlators involved, for instance, in problems of feed-back control
of the dynamics of the measured system. For measurement process
governed by Eq.\ (\ref{e5}) such a description is obtained by
first solving this equation in terms of $n$. Noticing that Eq.\
(\ref{e5}) coincides in essence with the recurrence relations for
the modified Bessel functions $I_n$ and assuming initial
condition $\rho^{n}_{kl}(0)= \rho_{kl}(0)\delta_{n,0}$ we get:
\begin{eqnarray}
\rho^{n}_{kl}(\tau)=\rho_{kl}(0)(\Gamma_+/\Gamma_-)^{n/2}
I_n (2\tau|t_kt_l|\sqrt{\Gamma_+\Gamma_-}) \cdot \nonumber \\
\exp \{ -(1/2)(\Gamma_+ +\Gamma_-) (|t_k|^2+|t_l|^2) \tau
-in \varphi_{kl} \} \, .
\label{e7} \end{eqnarray}
The qubit density matrix conditioned on the particular ``observed''
number $n$ of transferred particles is obtained then by selecting
the term with this $n$ in Eq.\ (\ref{e7}) and normalizing
the resulting reduced density matrix.

Quantitatively, conditional ``Bayesian'' equations for qubit
evolution can be derived starting from Eq.\ (\ref{e7}) and
following the same steps as in \onlinecite{ank}. In particular,
for weak detector-qubit coupling,
$|\delta_j|,|\lambda|\ll |t_0|$, when individual tunneling events
do not provide significant information on the qubit state, it is
convenient to condition the evolution on the quasicontinuous
current $I(t)$ in the detector. Then, conditional equation for
the evolution of the qubit density matrix is:
\begin{eqnarray}
&& \dot{\rho}_{kl} = -i[H_0,\rho]_{kl} -\gamma_{kl}\rho_{kl}
\nonumber \\
&& + I_f(t) \rho_{kl} \big(\frac{1}{2S_0}\sum_j
\rho_{jj}(I_k+I_l-2I_j)- i \varphi_{kl} \big)\, . \label{e9}
\end{eqnarray} where $S_0$ is the background current noise (see
Eq.\ (\ref{ea2})), variation of which with the qubit state can be
neglected in the weak-coupling limit, $S_0=(\Gamma_+
+\Gamma_-)|t_0|^2$. Also, $I_k= (\Gamma_+ -\Gamma_-) |t_k|^2$ is
the average detector current in the qubit state $k$, and $I_f(t)=
I(t)-\sum_k \rho_{kk} I_k$ is the fluctuation component of the
detector current. Equation (\ref{e9}) is written in the It\^o
form, in which averaging over $I_f(t)$ can be done by simply
omitting the terms with it. In the weak-coupling regime,
$\gamma_{kl}=(1/2) (\Gamma_+ +\Gamma_-)[(|t_k|-|t_l|)^2+
\varphi_{kl}^2|t_0|^2]$.
It is the same ensemble-averaged
decoherence rate as in Eq.\ (\ref{e6}), but in Eq.\ (\ref{e9}),
it leads to decoherence only after averaging over $I_f(t)$.

We now use equations obtained above to discuss several quantitative
characteristics of quadratic measurements. We start with the
purely quadratic detectors, when $\delta_j=0$, so that
$I_1=I_4=(\Gamma_+ -\Gamma_-) |t_0+\lambda|^2 \equiv I_{\uparrow
\uparrow}$ and $I_2=I_3=(\Gamma_+ -\Gamma_-) |t_0-\lambda|^2
\equiv I_{\uparrow \downarrow}$. In this case, if the qubits are
{\em stationary}, $H_0=0$, the detector effectively measures the
product operator $\sigma_z^{1} \sigma_z^{2}$ of the two qubits.
I.e., on the time scale of measurement time $\tau_m=4S_0/
(I_{\uparrow \uparrow}-I_{\uparrow \downarrow})^2$, the
subspace $\{|1\rangle,
|4\rangle\}$ in which the states of the two qubits are the same
and the average detector current is $I_{\uparrow \uparrow}$, is
distinguished from the subspace $\{|2\rangle,|3\rangle\}$ in
which the states of the two qubits are opposite and the current
is $I_{\uparrow \downarrow}$, while the states within these
subspaces are not distinguished. This property of quadratic
measurements can be used to design simple error-correction
scheme for dephasing errors \cite{err}.

Next, we consider the case of {\em identical, unbiased,
non-interacting} qubits with non-vanishing Hamiltonian,
$H_0= -(\Delta/2)\sum_j \sigma_{x}^{j}$. In this case the two
degenerate zero-energy eigenstates of $H_0$ can be chosen as
$\{|\!\uparrow \uparrow\rangle-|\!\downarrow\downarrow\rangle, \,
|\!\uparrow \downarrow\rangle-|\!\downarrow\uparrow \rangle\}$.
In the remaining subspace that will be denoted $D_+$, in the
basis $\{|\!\uparrow \uparrow \rangle+|\!\downarrow \downarrow
\rangle, \, |\!\uparrow \downarrow
\rangle+|\!\downarrow\uparrow \rangle\}$, $H_0$ reduces to
$-\Delta \sigma_x$ and mixes the states with similar and
opposite states of the two qubits. Accordingly, there are three
possible measurement outcomes characterized by the different
dc currents $\langle I\rangle$ in the detector, $I_{\uparrow
\uparrow}$, $I_{\uparrow \downarrow}$, and $(I_{\uparrow
\uparrow}+ I_{\uparrow \downarrow})/2$. These outcomes
can be interpreted as measurement of the operator $\sigma_y^1
\sigma_y^2 +\sigma_z^1\sigma_z^2$. Conditional equation
(\ref{e9}) can be used to simulate how the qubits, on the time
scale $\simeq 4S_0/I_a^2$, $I_a \equiv (I_{\uparrow \uparrow}-
I_{\uparrow \downarrow})/2$, are driven into one of the three
outcomes driven by the specific realization of the detector
current. The probabilities of
different outcomes depend on the initial qubit state. In the
first two outcomes, the initial state is projected on one of
the fully entangled states of the two qubits, e.g.,
\begin{equation}
\langle I\rangle= I_{\uparrow \downarrow} \;\;
\Leftrightarrow \;\; | \psi \rangle=(|\!\uparrow \downarrow
\rangle-|\!\downarrow \uparrow \rangle)/\sqrt{2}\, .
\label{e10} \end{equation}
Thus, quadratic measurements of two symmetric qubits provides
a simple way of generating entangled states of qubits that
in contrast to linear measurements \cite{lin} is based only
on monitoring dc current instead of spectrum.

In the third scenario, when $\langle I\rangle=
(I_{\uparrow \uparrow}+
I_{\uparrow \downarrow})/2$, the two qubits are confined to
the subspace $D_+$ and perform coherent quantum oscillations.
Equation for the density matrix (\ref{e6}) reduced to $D_+$
is:
\begin{equation}
\dot{\rho}_{kl} = i\Delta [\sigma_x,\rho]_{kl} -\gamma
\left( \begin{array}{cc} 0\, , & \rho_{12}  \\
\rho_{21}\, , & 0 \end{array} \right) ,
\label{e11} \end{equation}
where $\gamma \equiv 2 (\Gamma_+ +\Gamma_-)|\lambda|^2$.
Solving this equation and using the fact that the current
operator (\ref{ea1}) is reduced in $D_+$ to $I_a\sigma_z$,
we find the current spectral density (\ref{ea2}):
\begin{equation}
S_I(\omega ) = S_0 + \frac{8\Delta^2I_a^2 \gamma}
{(\omega^2-4\Delta^2)^2+\gamma^2\omega^2},
\label{e12} \end{equation}
where $S_0=(\Gamma_+ +\Gamma_-)(|t_0|^2+|\lambda|^2)$.
Qualitatively, for $\gamma \ll \Delta$, spectral density
(\ref{e12}) describes coherent oscillations of the two qubits
with the frequency $2\Delta$, twice the oscillation frequency
in one qubit. Similarly to the case of linear measurements
\cite{b6}, the maximum of the ratio of the oscillation peak
versus noise $S_0$ is 4. As one can see from Eq.\ (\ref{e12}),
this maximum is reached in the case of weak measurement
$|\lambda|\ll |t_0|$ by the ``ideal'' detector for which
$\arg(t_0\lambda^*)=0$ and only $\Gamma_+$ or $\Gamma_-$ is
non-vanishing.

For {\em different qubit tunnel amplitudes}, transitions (with
the rate $(\Delta_1 -\Delta_2)^2/2\gamma$ for small
$\Delta_1 -\Delta_2$) between the states
$|\!\uparrow \uparrow\rangle-|\!\downarrow\downarrow\rangle$ and
$|\!\uparrow \downarrow\rangle-|\!\downarrow\uparrow \rangle$
mix the measurement outcomes $I_{\uparrow \uparrow}$ and
$I_{\uparrow \downarrow}$. This means that for $\Delta_1 \neq
\Delta_2$ there are only two outcomes that have
the same dc current and differ by current spectral densities.
In one, the qubits are again in the subspace $D_+$ and the
spectral density is given
by Eq.\ (\ref{e12}) where now $2\Delta \rightarrow \Delta_1 +
\Delta_2$. In the other, the qubit dynamics is confined to
the subspace orthogonal to $D_+$, the basis of which can be
chosen as $\{|\!\uparrow \uparrow \rangle-|\!\downarrow
\downarrow \rangle, \, |\!\uparrow \downarrow \rangle-|
\!\downarrow\uparrow \rangle\}$. Evolution of the qubit
density matrix is described in this basis by the same Eq.\
(\ref{e11}) with $\Delta \rightarrow (\Delta_1 -\Delta_2)/2$,
and as a result the current spectral density
is given by the same Eq.\ (\ref{e12}) with $2\Delta
\rightarrow \Delta_1 -\Delta_2$ and describes the current
peak at difference of qubit oscillation frequencies.

As the last application of the general theory we consider
two identical qubits measured by a weakly and symmetrically
coupled detector with {\em arbitrary non-linearity}. It is
convenient to discuss this situation in terms of the total
effective spin $S$ of the two qubits which determines the
amplitude (\ref{e4}) of detector tunneling:
\begin{equation}
t=t_0+2 \delta S_z + \lambda (2 S_z^2-1)\, .
\label{e13} \end{equation}
The $S=0$ state (\ref{e10}) does not evolve in time under
the qubit Hamiltonian and represents one of the
measurement outcomes characterized by the dc detector
current $I_{\uparrow \downarrow}$ and flat current spectral
density $S_I(\omega)=(\Gamma_+ +\Gamma_-)|t_0-\lambda|^2$.
Three other, $S=1$, states are mixed by measurement and
represent the second measurement outcome. We take the basis
of the  $S=1$ subspace as three energy eigenstates
$S_x=-1,0,1$ with energies $\{\Delta,0,-\Delta\}$. Interaction
with the detector induces transitions between these states
with the rate independent of the transition's direction, so
that the stationary qubit density matrix is $\rho_0=1/3$.
Equations (\ref{ea1}) and (\ref{e13}) show then that the
dc detector current for this outcome is:
\[ \langle I\rangle = (\Gamma_+ -\Gamma_-)[2
(|t_0|^2+|\lambda|^2)+|t_0+\lambda|^2+8|\delta|^2]/3 \, , \]
and can be written as $\langle I\rangle = (I_{\uparrow \uparrow}
+ I_{\downarrow \downarrow}+ I_{\uparrow \downarrow})/3$, where
the currents $I_{\uparrow \uparrow}, ...$ are introduced in the
same way as before, e.g, $I_{\uparrow \uparrow}=
(\Gamma_+ -\Gamma_-)|t_0+2\delta+\lambda|^ 2$. The background
current noise $S_0$ coincides with $\langle I \rangle$ with
$\Gamma_+ -\Gamma_-$ replaced by $\Gamma_+ +\Gamma_-$.

The system performs oscillations at frequencies $\Delta$ and
$2\Delta$ whose spectral peaks should have Lorentzian form for
weak
detector-qubit coupling. Evaluating the current matrix elements
from Eqs.\ (\ref{ea1}) and (\ref{e13}), and evolution of the
density matrix from Eq.\ (\ref{e6}) reduced to the $S=1$ subspace,
we obtain parameters of these Lorentzians in the spectral density
(\ref{ea2}):
\begin{equation}
\omega \simeq j\Delta\, , \;\;\; S_I(\omega)=S_0+\frac{2}{3}
\frac{a_j^2 \gamma_j}{(\omega-j\Delta)^2 +\gamma_j^2} \; ,
\label{e15}  \end{equation}

\vspace*{-3ex}

\begin{eqnarray}
j=1,2 \, , \;\;\; \gamma_1 = (\Gamma_+ +\Gamma_-)
(|\delta|^2 +|\lambda|^2/2) \, , \;\; \gamma_2=2\gamma_1 \, ,
\nonumber \\
a_1=4(\Gamma_+ -\Gamma_-) \mbox{Re} [(t_0+\lambda)\delta^*] =
(I_{\uparrow \uparrow}-I_{\downarrow \downarrow})/2\, ,
\nonumber \\
a_2=2(\Gamma_+ -\Gamma_-) (\mbox{Re} [t_0\lambda^*]+|\delta|^2)
= (I_{\uparrow \uparrow}+I_{\downarrow \downarrow}-2
I_{\uparrow \downarrow})/4\, . \nonumber
\end{eqnarray}
Note that condition $\rho_0=1/3$, used in Eq.\ (\ref{e15}) and
Fig.\ 2, implies that the coefficient $\delta$ of linear
measurement mixing all three $S=1$ states, does not vanish
identically. Otherwise, only two states are mixed and
$\rho_0=1/2$, as assumed in Eq.\ (\ref{e12}). In general, there
is also a spectral peak at $\omega =0$ caused by switching
between states with different average currents. For small but
finite $\delta$, $\delta \ll \lambda$, this peak can be very
high, e.g., in the case of an ``ideal'' detector ($\Gamma_-=\arg
t_0\lambda^*= \arg t_0\delta^*=0$) its height and half-width are,
respectively, $(8\lambda^2/27\delta^2) S_0$ and $6\delta^2
\Gamma_+$.

Figure 2 shows current spectral density calculated from Eqs.\
(\ref{e6}) and (\ref{ea2}) for an ideal detector without making
weak-coupling approximation. Figure 2a illustrates the transition
between ``single-qubit'' oscillations at $\omega \simeq \Delta$
in the case of linear measurement and oscillations at $\omega
\simeq 2\Delta$ for quadratic measurement. One can see that in
agreement with Eq.\ (\ref{e15}) the $\omega \simeq \Delta$ peak
is typically higher than the quadratic peak at double frequency.
It is at the same time more sensitive to qubit-qubit interaction
as illustrated in Fig.\ 2b. Even weak interaction $\nu \ll
\Delta$ first suppresses and then splits the peak at $\omega
\simeq \Delta$ in two while affecting the quadratic peak only
slightly.

In summary, we discussed quadratic quantum measurements and have
shown that they should have non-trivial properties providing, for
instance, simple method of entangling non-interacting qubits. We
also calculated output spectra of the quadratic detector
measuring coherent oscillations in two qubits. Consistent with
the case of classical oscillations, quadratic measurements
results in the spectral peak at frequency that is twice the
frequency of individual qubit oscillations. Quadratic
measurements should be an interesting and potentially useful tool
in solid-state quantum devices.

\begin{figure}[h]
\setlength{\unitlength}{1.0in}
\begin{picture}(3.0,3.45)
\put(.1,-0.08){\epsfxsize=2.8in\epsfbox{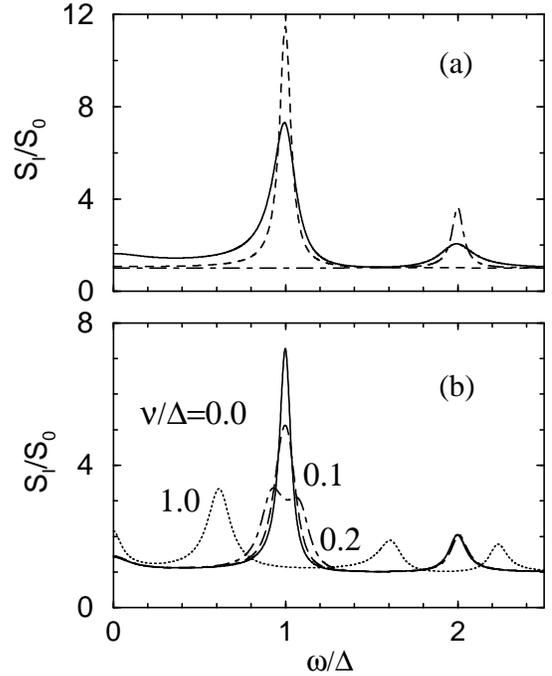}}
\end{picture}
\caption{Spectra of coherent quantum oscillation in two qubits
measured with arbitrary non-linearity. (a) Transition from linear
to quadratic measurement in the case of non-interacting qubits,
$\nu=0$, for $\Gamma_+ |t_0|^2=4.0\Delta$. Dashed line, solid
line and dot-dashed line correspond to linear, ($\delta=0.1t_0$,
$\lambda=0$), intermediate ($\delta=\lambda= 0.1t_0$), and
quadratic ($\delta=0$, $\lambda=0.1t_0$) case. $\;$ (b) Effect of
qubit-qubit interaction on spectrum for $\Gamma_+
|t_0|^2=2.0\Delta$ and $\delta=\lambda=0.1t_0$.}
\end{figure}

\vspace*{.4ex}

This work was supported in part by ARDA and DOD under the DURINT
grant \# F49620-01-1-0439 and by the NSF under grant \# 0121428
(W.M. and D.V.A.), and also by NSA and ARDA under the ARO grant
DAAD19-01-1-0491 (R.R. and A.N.K.).


\begin{references}
\bibitem{b5} In some regimes, the particle transfer through the
detector takes place in more than one step and can not be
characterized by one transmission amplitude $t$. In this work,
we concentrate on the case of one-step transfer, since only in
this case the detector can be quantum-limited.
\end{references}
\end{document}